\documentclass[a4paper,11pt]{article}
\pdfoutput=1 
\usepackage{jheppub, bm, color} 
\usepackage[T1]{fontenc}
\usepackage{amssymb,amsfonts,slashed,amsthm,amsmath,graphicx}
\preprint{
IPMU 17-0114;
TU-1046
}

\title{
Adiabatic suppression of the axion abundance and isocurvature due to coupling to hidden monopoles
}

\author[1,2]{
Masahiro Kawasaki
}
\author[3,2]{
Fuminobu Takahashi
}
\author[4,3]{
Masaki Yamada
}

\affiliation[1]{Institute for Cosmic Ray Research, 
The University of Tokyo, 
Kashiwa, Chiba 277-8582, Japan}
\affiliation[2]{Kavli IPMU (WPI), UTIAS, 
The University of Tokyo, 
Kashiwa, Chiba 277-8583, Japan}
\affiliation[3]{Department of Physics, Tohoku University, 
Sendai, Miyagi 980-8578, Japan} 
\affiliation[4]{Institute of Cosmology, Department of Physics and Astronomy, 
Tufts University, Medford, MA  02155, USA}

\newcommand{\eV}{ \ {\rm eV} }

\newcommand{\MeV}{\  {\rm MeV} }
\newcommand{\GeV}{\  {\rm GeV} }
\newcommand{\TeV}{\  {\rm TeV} }
\newcommand{\PeV}{\  {\rm PeV} }

\newcommand{\lmk}{\left(}  
\newcommand{\rmk}{\right)}
\newcommand{\lkk}{\left[}  
\newcommand{\rkk}{\right]}

\newcommand{\del}{\partial}  
\newcommand{\la}{\left\langle} 
\newcommand{\ra}{\right\rangle}

\newcommand{\bea}{\begin{array}}
\newcommand{\eea}{\end{array}}
\newcommand{\beq}{\begin{eqnarray}}
\newcommand{\eeq}{\end{eqnarray}}

\newcommand{\dd}{\mathrm{d}}
\newcommand{\Mpl}{M_{\rm Pl}}

\newcommand{\abs}[1]{\left\vert {#1} \right\vert}

\newcommand{\Max}{\text{Max}}

\newcommand{\cphi}{\varphi}

\newcommand{\Lqcd}{\Lambda_{\rm QCD}}
\newcommand{\const}{\text{const}}

\def\lrf#1#2{ \left(\frac{#1}{#2}\right)}
\def\lrfp#1#2#3{ \left(\frac{#1}{#2} \right)^{#3}}

\def\EQ#1{Eq.~(\ref{#1})}

\def\GEV#1{10^{#1}{\rm\,GeV}}

\newcommand{\fa}{f_{a}}

\newcommand{\mat}{m_{a}^2 (T)}
\newcommand{\mam}{m_{a, {\rm M}}^2 }
\def\eq#1{Eq.~(\ref{#1})}

\newcommand{\cm}{\ {\rm cm}}
\newcommand{\km}{\ {\rm km}}

\newcommand{\LQCD}{\Lambda_{\rm QCD}}

\abstract{
The string theory predicts many light fields called moduli and axions, which cause a cosmological problem due to the overproduction of their coherent oscillation after inflation. One of the prominent solutions is an adiabatic suppression mechanism, which, however, is non-trivial to achieve in the case of axions because it necessitates a large effective mass term which decreases as a function of time. The purpose of this paper is twofold. First, we provide an analytic method to calculate the cosmological abundance of coherent oscillation in a general situation under the adiabatic suppression mechanism. Secondly, we apply our method to some concrete examples, including the one where a string axion acquires a large effective mass due to the Witten effect in the presence of hidden monopoles.
}

\begin{document}
\maketitle
\flushbottom

\section{Introduction
\label{sec:introduction}}

The string theory is a candidate for the theory of everything. 
While it is hard to derive the Standard Model (SM) of particle physics in a top-down manner,  it provides intriguing insight into the low energy physics. One of the important implications is the existence of many light fields called moduli and axions~\cite{Witten:1984dg, Svrcek:2006yi}.  They acquire a mass from supersymmetry (SUSY) breaking effects and other non-perturbative effects. Since their interactions with the SM particles are typically suppressed by the Planck scale, those light fields are long-lived and play a major role in cosmology. In fact, they are known to cause a catastrophic problem for the evolution of the Universe. 

During inflation, those light fields may be displaced from the low-energy minima, since their potentials are very flat and in some case, the potential could be significantly modified by the SUSY breaking effects of the inflaton. 
After inflation ends, the energy scale of the Universe decreases and eventually the Hubble parameter becomes as low as the mass scale of the light fields. Then the light fields start to oscillate coherently around the minimum of the potential~\cite{Preskill:1982cy}. The amplitude of the oscillation is expected to be of order the Planck scale or the axion decay constant. The energy density of the coherent oscillation behaves like matter and decreases slower than radiation so that they dominate the Universe soon after the onset of oscillation. Since the interactions with the SM particles are suppressed by the Planck scale, their lifetime may be longer than the big bang nucleosynthesis (BBN) epoch and the resulting evolution of the Universe would not be consistent with observations. 
This is the notorious cosmological moduli problem.%
\footnote{
Another important prediction of the string theory in cosmology is the string landscape. 
There are an exponentially large number of vacua in the field space of moduli and string axions 
and 
the anthropic selection of vacua may explain the fine tuning of the cosmological constant. 
The slow-roll inflation in the string landscape is also widely studied 
(see, e.g., Ref.~\cite{Baumann:2014nda} 
for a review 
and 
Refs.~\cite{
Bachlechner:2014rqa,Higaki:2014pja,Higaki:2014mwa,Wang:2015rel,Masoumi:2016eqo,Dias:2016slx,Wang:2016kzp,Daido:2016tsj,Freivogel:2016kxc,Pedro:2016sli,Masoumi:2016eag,Easther:2016ire,Masoumi:2017gmh,Dias:2017gva,Masoumi:2017xbe} 
for recent works.) 
}

A novel solution to the problem was suggested by Linde about two decades ago~\cite{Linde:1996cx}. 
If the light fields obtain a time-dependent effective mass much larger than the Hubble parameter, 
they may follow  the slowly-moving potential minimum adiabatically. As a result, the oscillation energy is
 suppressed when the Hubble parameter becomes comparable to their low-energy mass.\footnote{
Thermal inflation is another possible solution 
to the cosmological moduli problem, 
where the moduli density is diluted by a mini-inflation 
around the TeV scale~\cite{Yamamoto:1985rd, Lyth:1995ka,  Asaka:1997rv, Asaka:1999xd}. 
See Refs.~\cite{Hayakawa:2015fga, Arai:2016vwa} for recent works. 
}
In fact, however, the suppression of the moduli abundance
is not so efficient as originally expected~\cite{Nakayama:2011wqa,Nakayama:2011zy},
since some amount of moduli oscillations is necessarily induced after inflation. This is because the inflaton is lighter than the Hubble parameter
for slow-roll inflation and later becomes much heavier than the Hubble parameter and the moduli mass.  
The oscillation of the inflaton becomes relevant for the moduli dynamics especially when their mass scales 
become comparable.
Also, we note that it is non-trivial to realize the adiabatic suppression mechanism 
for axions, because the axion potential is protected by a shift symmetry and is induced
from non-perturbative effects.

Recently, we found in Ref.~\cite{Kawasaki:2015lpf} 
that the adiabatic suppression mechanism works for axions coupled to 
a hidden U(1)$_H$ gauge symmetry with a hidden monopole. 
In the presence of hidden monopoles, 
the theta parameter in the hidden U(1)$_H$ has a physical effect, known as the Witten effect, 
where the monopole acquires a hidden electric charge proportional to the theta parameter~\cite{Witten:1979ey}. 
Here, when the Peccei-Quinn (PQ) symmetry is also anomalous under the hidden U(1)$_H$ symmetry, 
the theta parameter of the hidden U(1)$_H$ is promoted to the dynamical axion field~\cite{Fischler:1983sc}. 
This implies that the hidden monopole acquires an electric charge related to the axion field value. 
Since the nonzero value of electric charge is not favored 
to minimize the total energy, 
the axion starts to cancel the theta parameter of the hidden U(1)$_H$ 
to make 
the hidden electric charge of monopole absent. 
This means that the axion acquires an effective mass via the Witten effect and its abundance can be suppressed by the adiabatic suppression mechanism. 
Also, the axion isocurvature can be suppressed by the Witten effect~\cite{Kawasaki:2015lpf, Nomura:2015xil}. 
See e.g. Refs.~\cite{Linde:1990yj,Linde:1991km,Lyth:1992tw,Kasuya:1996ns,Dine:2004cq,
Folkerts:2013tua,Jeong:2013xta,Higaki:2014ooa,Dine:2014gba,Nakayama:2015pba,Harigaya:2015hha,
Choi:2015zra,Kawasaki:2015lea,Takahashi:2015waa,Agrawal:2017eqm} for other scenarios to suppress the axion isocurvature.

The purpose of the present paper is twofold. 
First, we provide an analytic method to calculate the abundance of light fields under the adiabatic suppression mechanism. 
We calculate an approximately conserved adiabatic invariant, which represents the comoving number density of light fields, and show that the adiabatic invariant is nonzero but exponentially suppressed if we neglect the initial abundance which may be induced by the effect of the inflaton oscillation. 
Our result is consistent with Ref.~\cite{Linde:1996cx}, but our method can be applied to a more generic situation. 
Secondly, we apply our calculation to some concrete models, including the one with the Witten effect on the string axion.

This paper is organized as follows. 
In the next section, 
we briefly review the moduli problem and the axion overproduction problem. 
Then in Sec.~\ref{sec:calculation}, we calculate the abundance of coherent oscillation of scalar field under the adiabatic suppression mechanism. Then in the subsequent sections, we apply the calculation to scenarios where an axion acquires an effective mass due to the Witten effect in the presence of monopoles in a hidden sector. 
In Sec.~\ref{sec:QCD axion}, 
we focus on the QCD axion and discuss a possibility that the dark matter (DM) is a hidden monopole. 
In Sec.~\ref{sec:string axion}, 
we show that the overproduction problem of string axion can be ameliorated by the Witten effect when abundant monopoles disappear before the BBN epoch. 
Finally, we conclude in Sec.~\ref{sec:conclusion}.

\section{Overproduction problem of light scalar fields
\label{sec:problem}}

The string theory predicts many moduli and axions that have extremely flat potentials 
compared with the fundamental scale, such as the Planck scale~\cite{Witten:1984dg, Svrcek:2006yi}. 
Such light fields start to oscillate around the low energy vacuum 
after inflation. 
Since the amplitude of the coherent oscillation is expected to be of order 
the Planck or GUT scale, 
they soon dominate the Universe, and the subsequent evolution of the Universe is 
inconsistent with that of our Universe. 
In this section, we briefly explain the overproduction problem of moduli and axions.

\subsection{Moduli problem}

The string theory predicts many light singlet scalar fields 
in low energy effective theory. 
Its potential is generically written as 
\beq
 V (\phi) = \frac{1}{2} m_\phi^2 \phi^2, 
\eeq
around the low-energy vacuum, 
where $m_\phi$ is the mass of moduli that is related to the SUSY breaking scale. 
Here we take $\phi = 0$ at the low energy vacuum by shifting its field value. 

The mass of moduli is expected to be in the range between TeV scale 
and PeV scale in most SUSY models 
motivated by addressing the hierarchy problem and the gauge coupling unification. 
In this paper, we consider the case where the 
Hubble parameter during inflation is much larger than 
the mass of moduli. 
In this case, 
the potential of the moduli is so flat that 
it may have a nonzero vacuum expectation value (VEV) during inflation because of the Hubble friction effect. 
Since the coherent length is stretched by the exponential expansion of the Universe during inflation, 
the moduli VEV becomes spatially homogeneous over the observable Universe. 
After inflation ends, the Hubble parameter $H(t)$ decreases and the moduli start to oscillate around the low-energy vacuum 
at $H(t) \simeq m_\phi$. 
The amplitude of its oscillation is expected to be of order 
the Planck scale, 
so that the resulting abundance is given by 
\beq
 \frac{\rho_\phi}{s} \simeq 
 \frac{1}{8} T_{\rm RH}, 
\eeq
where $T_{\rm RH}$ is reheat temperature after inflation. 
This energy density is much larger than the DM abundance. 
If the moduli is stable, the Universe would be in matter domination before the BBN starts. 
Even if the moduli is unstable, its decay products may spoil the success of the BBN~\cite{Kawasaki:2004qu}. 
This is the notorious cosmological moduli problem. 
Also, if the moduli is stabilized by SUSY breaking effects in the K\"ahler potential, their axionic partners remain
much lighter than the moduli. In this case, the moduli decay into a pair of axions with a large branching
ratio, leading to the overproduction of axions~\cite{Cicoli:2012aq,Higaki:2012ar,Higaki:2013lra}. Similarly,
the moduli generically decay into a pair of gravitinos with a sizable branching fraction if kinematically 
accessible~\cite{Endo:2006zj,Nakamura:2006uc,Dine:2006ii,Endo:2006tf}. Thus produced gravitinos may
spoil the BBN or produce too many lightest SUSY particles.

\subsection{Axion overproduction problem}

The potential of axion can be similarly written as follows around 
the low-energy vacuum: 
\beq
 V(a) \approx \frac{1}{2} m_{a}^2 a^2, 
 \label{axion potential QCD}
\eeq
where $a$ is the axion field and $m_{a}$ is the axion mass at the potential minimum.

For the QCD axion, 
its mass is given as 
\beq
 \left. m_{a} \right\vert_{T=0} \simeq 
 \frac{z}{(1 + z)^2} \frac{m_\pi f_\pi}{\fa}, 
\eeq
at the zero temperature, 
where $f_a$ is the axion decay constant, 
$z$ ($\simeq 0.56$) is the ratio of $u$- and $d$-quark masses, 
$m_\pi$ ($\simeq 140 \MeV$) is the pion mass, 
and $f_\pi$ ($\simeq 130 \MeV$) is the pion decay constant. 
In a finite temperature plasma 
with $T \gg \LQCD$, 
the axion mass depends on the temperature as
\beq
 \mat 
 \simeq 
 c_T \frac{\Lqcd^4}{\fa^2} 
 \lrfp{T}{ \Lqcd }{-n}, 
 \label{axion mass at T}
\eeq
where $c_T \simeq 1.68 \times 10^{-7}$, $n=6.68$, and $\Lqcd = 400 \MeV$~\cite{Kim:2008hd,Wantz:2009it}.

Let us consider a case in which the PQ symmetry is broken before inflation. 
In this case, 
the axion stays at a certain phase during inflation 
because it is massless and is affected by the Hubble friction effect. 
Since the coherent length is stretched by the exponential expansion of the Universe during inflation, 
the axion VEV is spatially homogeneous. 
After inflation ends and before the QCD phase transition, 
the axion mass is much smaller than the Hubble parameter, 
so that it continues to stay at a certain VEV due to the Hubble friction effect. 
Then at a time around the QCD phase transition, 
the axion mass becomes larger than the Hubble parameter 
and starts to oscillate around the low energy vacuum, 
at which the axion VEV cancels the undesirable strong CP phase~\cite{Preskill:1982cy}. 
The temperature of the Universe at the onset of oscillation is given as 
\beq
 T_{\rm osc, 0} 
 &\simeq& 
 \Lqcd 
 \lrfp{90 c_T \Mpl^2}{\pi^2 g_* (T_{\rm osc, 0}) \fa^2}{1/(4+n)} 
 \\
 &\simeq& 1.2 \GeV \lrfp{\fa}{10^{12} \GeV}{-0.187}, 
\eeq
where $\Mpl$ ($\simeq 2.4 \times 10^{18} \GeV$) is the reduced Planck mass. 
The parameter $g_* (T)$ is the effective number of relativistic particles in the plasma 
and we use $g_* (T_{\rm osc, 0}) \approx 85$. 
The energy density of the axion oscillation 
decreases with time 
just like matter after the axion mass becomes constant, 
and hence it is a good candidate for cold DM. 
Neglecting the anharmonic effect~\cite{Kobayashi:2013nva}, 
one obtains the axion abundance as~\cite{Bae:2008ue}
\beq
\Omega_a h^2 \;\simeq\; 0.2\, \theta_{\rm ini}^2 \lrfp{f_a}{\GEV{12}}{1.19}, 
\label{Omegaa}
\eeq
where $h$ is the Hubble parameter in units of $100 \ {\rm km/s/Mpc}$ 
and $\theta_{\rm ini}$ ($-\pi < \theta_{\rm ini} \le \pi$) is the initial misalignment angle. 
The observed DM abundance, $\Omega_{\rm DM} h^2 \simeq 0.12$, can be explained when the axion decay constant is given by 
\beq
 f_a \simeq 7.4 \times 10^{11} \GeV \times \abs{\theta_{\rm ini}}^{-1.68}. 
 \label{f_a for DM}
\eeq

Note that the string theory 
predicts axions with the axion decay constants of order the grand-unified theory (GUT) scale. 
With such a large decay constant, 
the axion abundance easily overcloses the Universe unless the initial misalignment angle $\theta_{\rm ini}$ is fine-tuned to be much smaller than unity 
[see \eq{Omegaa}]. 
Although the string axion is well motivated, 
it confronts the overproduction problem of the axion energy density.

\section{Adiabatic suppression mechanism
\label{sec:calculation}}

The moduli and axion overproduction problems can be avoided 
when the VEV of these fields changes adiabatically at the onset of oscillation. 
This can be achieved when they have a large time-dependent mass term in addition to the 
low-energy mass term~\cite{Linde:1996cx}. 
In this section, we explain the adiabatic suppression mechanism 
and show that it results in the exponential suppression of an adiabatic invariant, which describes the comoving number density of particles.

Suppose that 
a light field (moduli or axion) has a time-dependent mass $\tilde{m}(t)$. 
The potential of the light field $\phi$ is then given by 
\beq
 V(\phi) = \frac{1}{2} m_\phi^2 (t) \phi^2 
 + \frac{1}{2} \tilde{m}^2 (t) \lmk \phi - \phi_0 \rmk^2, 
\eeq
where $\phi_0$ is of order the Planck scale for the case of moduli 
or of order the axion decay constant for the case of axion. 
We denote the mass of light field as $m_\phi(t)$, which may depend on 
time in some cases. 
Note that the model considered in Ref.~\cite{Linde:1996cx} 
corresponds to the case with $m_\phi (t) = {\rm const.}$ ($\equiv m_\phi$) 
and $\tilde{m}^2 (t) = C^2 H^2(t)$ ($C$: constant). 
The equation of motion is then given by 
\beq
 \ddot{\phi} + 3 H \dot{\phi} = - m_\phi^2 (t) \phi - \tilde{m}^2(t) \lmk \phi - \phi_0 \rmk, 
 \label{axion EOM}
\eeq
where $H = p/t$ with $p=1/2$ in the radiation dominated era.

We assume that $\tilde{m} (t)$ is larger than $H(t)$ 
at least until $\tilde{m} (t) \simeq m_\phi (t)$. 
In this case $\phi$ stays at or oscillates around $\phi = \phi_0$. 
Then, at the time around $\tilde{m}(t) \simeq m_\phi (t)$, 
the minimum of the potential decreases as 
$v (t) \equiv \phi_0 \tilde{m}^2 (t) / (m_\phi (t) ^2 + \tilde{m}^2 (t))$. 
Here, the time scale of $v (t)$ is of order the Hubble parameter ($\dot{v} / v \sim H(t)$) 
while that of the oscillation is of order $\tilde{m}(t)$ ($ \gg H(t)$). 
This means that the minimum of the potential changes adiabatically 
and almost no oscillation is induced through this dynamics. 

Note that we should take a particular care of the origin of the effective mass term. Suppose that, for example, the effective mass term comes from a coupling between the moduli field and the inflaton. During inflation, the inflaton mass is lighter than the Hubble parameter for successful slow-roll inflation. After inflation, the inflaton oscillates about its potential minimum and its mass becomes larger than the Hubble parameter. This implies that the effective modulus mass becomes necessarily comparable to the inflaton mass sometime after inflation.  Then, the modulus dynamics can be significantly affected by the inflaton oscillations, leading to a production of the modulus field after inflation~\cite{Nakayama:2011wqa}. 
The model used in Ref.~\cite{Linde:1996cx} confronts this issue. On the other hand, the examples we use in Sec.~\ref{sec:QCD axion} and \ref{sec:string axion} are free from this issue, since the effective axion mass appears after inflation when the inflaton mass is already much heavier
than the effective axion mass.

Below, we evaluate the abundance of the light field by calculating the time evolution of an 
adiabatic invariant. 
Our calculation not only reproduces the result of Ref.~\cite{Linde:1996cx}
but also is applicable to more generic cases.

\subsection{Calculation of adiabatic invariant}

We rewrite the equation of motion as 
\beq
 \ddot{\cphi} + m^2 (t) \lmk \cphi - v (t) \rmk = 0, 
\eeq
where 
\beq
 \cphi &\equiv& \lmk \frac{t}{t_0} \rmk^{3p/2} \phi, 
 \label{a to cphi}
 \\
 m^2 (t) &\equiv& m_\phi^2(t) + \frac{3}{2} \lmk \frac{1}{p} - \frac{3}{2} \rmk H^2(t) 
 + \tilde{m}^2(t) 
 \\
 v(t) &\equiv& \frac{\tilde{m}^2 (t)}{m^2(t)} \phi_0 \lmk \frac{t}{t_0} \rmk^{3p/2}. 
\eeq
The dynamics of such a field is analogous to 
a motion of a particle 
with the Hamiltonian, 
\beq
 \mathcal{H} = \frac{1}{2} \Pi^2 + \frac{1}{2} m^2(t) \lmk \cphi - v(t) \rmk^2. 
\eeq
where $\Pi$ is a canonical momentum of the particle $\cphi$. 
This is the Hamiltonian of a harmonic oscillator with two time-dependent parameters.

We can define an adiabatic invariant 
for a one-particle system with a compact trajectory. 
We assume that 
the typical time scale of its oscillation $m(t)$ 
is much larger than $H(t)$ and 
those of the two parameters $(\dot{m}/m)^{-1}$ and $(\dot{v}/v)^{-1}$ are 
of order $H(t)$. 
In this case, 
the adiabatic invariant approximately conserves. 
It is explicitly written as 
\beq
 I = \frac{1}{2 \pi } \int_T \Pi \dd \cphi, 
\eeq
where the integral is taken over the interval of one periodic motion assuming constant $m$ and $v$. 
In an oscillating homogeneous scalar field, 
the adiabatic invariant can be interpreted as the comoving number density of particles, 
and its approximate conservation law 
means that almost no 
particles are produced due to slowly changing parameters. 
However, 
exponentially suppressed but nonzero amount of particles are produced throughout the dynamics 
and 
we can calculate it in the following way. 
We follow and generalize a method explained in Ref.~\cite{Landau}.

\subsubsection{Case with constant parameters}

First, let us consider a trivial case where the parameters $m$ and $v$ are constant in time. 
The result is of course given by 
\beq
 \cphi (t) &=& \sqrt{\frac{2 E}{m^2}} \sin \theta (t) + v, 
 \label{cphit}
 \\
 \Pi (t) &=& \sqrt{2 E} \cos \theta (t), 
 \label{Pit}
\eeq
where $E$ is the energy, $\Pi = \dot{\cphi}$ is the canonical momentum, 
and $\theta (t)$ is given by $m t$. 
The adiabatic invariant, which is exactly conserved in this case, is calculated as 
\beq
 I &=& E / m. 
\eeq
Note that since we define $\cphi$ as \eq{a to cphi}, 
the adiabatic invariant is proportional to the comoving number density of 
the field $\phi$.

Let us take the action as a function of $\cphi$ and $t$ such as 
\beq
 S(\cphi, t) = \int^{\cphi, t} \mathcal{L} \dd t'. 
\eeq
Noting that $\mathcal{L} = \Pi (\del \cphi / \del t) - \mathcal{H}$, 
where $\mathcal{H}$ is the Hamiltonian, 
we obtain 
\beq
 \dd S = \Pi \dd \cphi - \mathcal{H} \dd t. 
\eeq
Note that $\Pi$ should be rewritten in terms of $\cphi$ by using Eqs.~(\ref{cphit}) and (\ref{Pit}). 
Since the energy is conserved for the case of constant $m$ and $v$, 
it is given by 
\beq
 S (\cphi, t) &=& S_0 (\cphi, E) - E t,  \\
 S_0 (\cphi, E) &=& \int \Pi (\cphi', E) \dd \cphi', 
 \label{S_0}
\eeq
where we explicitly write $E$ dependence of $S_0$ for later convenience. 
Here, we can take $S_0(\cphi, E)$ as a function of $I$ because $E = m I$. 
In the case of the harmonic oscillator, 
$S_0$ is explicitly calculated as 
\beq
 S_0 (\cphi, I) 
 = \sqrt{2 m I} \int \sqrt{ 1 - \frac{m}{2 I} \lmk \cphi' - v \rmk^2 } \dd \cphi' 
 = I \lmk \theta + \frac{\sin 2 \theta}{2} \rmk, 
 \label{S_0 for H.O.}
\eeq
where $\theta = \theta (\cphi, I)$ is given by the inverse of \eq{cphit}.

Let us take $S_0$ as a mother function 
of a canonical transformation: 
\beq
 \frac{\dd S_0}{\dd t} (\cphi, I) = \lmk \Pi \frac{\dd \cphi}{\dd t} - \mathcal{H} \rmk 
 - \lmk I \frac{\dd \theta}{\dd t}  - \mathcal{H}' \rmk + \frac{\dd }{\dd t} \lmk I \theta \rmk, 
\eeq
where $\mathcal{H}'$ is a new Hamiltonian. The last terms comes from the Legendre transformation 
that change the dependence of the variable $\theta$ to $I$. 
Then the adiabatic invariant $I$ becomes a new canonical momentum 
and satisfies 
\beq
 \Pi &=& \frac{\del S_0 (\cphi, I)}{\del \cphi}, 
 \label{Pi}
\\
 \theta &=& \frac{\del S_0(\cphi, I)}{\del I}. 
 \label{theta}
\eeq
The parameter $\theta$ is a new canonical variable. 
It coincides with the one used in \eq{cphit}, as expected. 
Since the mother function $S_0$ is independent of time, 
the new Hamiltonian is identified with the old one, i.e., $\mathcal{H} = E(I)$. 
The Hamilton equations of motion are thus given by 
\beq
 \dot{I} = - \frac{\del E(I)}{\del \theta} = 0, 
\\
 \dot{\theta} = \frac{\del E(I)}{\del I} = m, 
\eeq
which 
imply that $I = {\rm const.}$ and $\theta = m t$ 
as expected.

\subsubsection{Case with time-dependent parameters}

Next, we consider a case where the parameters $m$ and $v$ depend on time. 
We perform a canonical transformation of this system 
via the mother function of $S_0$ given in Eq.~(\ref{S_0 for H.O.}), 
where the constant parameters $m$ and $v$ are replaced by the time-dependent ones. 
Since the mother function depends on time via the parameters $m(t)$ and $v(t)$, 
the new Hamiltonian is given by 
\beq
 \mathcal{H}' 
 = E(I) + \frac{\del S_0}{\del t}
 = E(I) + \Lambda_m \dot{m} + \Lambda_v \dot{v}, 
\eeq
where 
\beq
 \Lambda_m &=& \lmk \frac{\del S_0}{\del m} \rmk_{\cphi, I} = \frac{I}{2 m} \sin 2 \theta, 
\\
 \Lambda_v &=& \lmk \frac{\del S_0}{\del v} \rmk_{\cphi, I} = - \sqrt{2 m I} \cos \theta, 
\eeq
where $(\del f (x,y) / \del x)_y$ is the partial derivative of $f (x,y)$ with respect to $x$ 
while $y$ fixed.

The new canonical variables $I$ and $\theta$ are determined by Eqs.~(\ref{Pi}) and (\ref{theta}). 
The equations of motion are now given by 
\beq
 \dot{I} &=& 
 - \frac{\del \mathcal{H}'}{\del \theta}
 =  
 - \lmk \frac{\del \Lambda_m}{\del \theta} \rmk_{I, m, v} \dot{m} 
 - \lmk \frac{\del \Lambda_v}{\del \theta} \rmk_{I, m, v} \dot{v}, 
\\
 \dot{\theta} &=& 
  \frac{\del \mathcal{H}'}{\del I}
 =  
 \lmk \frac{\del E}{\del I} \rmk_{m, v} 
 + \lmk \frac{\del \Lambda_m}{\del I} \rmk_{\theta, m, v} \dot{m} 
 + \lmk \frac{\del \Lambda_v}{\del I} \rmk_{\theta, m, v} \dot{v}, 
 \label{theta dot}
\eeq
where $(\del E / \del I)_{m, v}$ ($= m$) is the frequency of the harmonic oscillator 
for the case of time-independent parameters. 
These are rewritten as 
\beq
 \dot{I}(t) &=& 
 -  I \frac{\dot{m}}{m}  \cos \lmk 2 \theta \rmk
 - \sqrt{2 m I} \dot{v} \sin \theta, 
 \label{dot I}
 \\
 \dot{\theta} &\simeq& m, 
\eeq
where we neglect the second and third terms in Eq.~(\ref{theta dot}) 
because the inverse of the oscillation time scale $m$ is much larger than 
that of $\dot{m} / m$ and $\dot{v} / v$.

On some occasions, we are interested in the case where 
the adiabatic invariant $I$ is absent initially. 
Then 
the second term of the right-hand side in Eq.~(\ref{dot I}) mainly contributes to the growth of $I$. 
Even if $I$ is initially nonzero, 
we can neglect the first term in the right-hand side 
for the case of $I \ll ( m / \dot{m} )^2 (\dot{v} / v)^2 m v^2  \approx m v^2$. 
For example, suppose that $\phi$ oscillates at $t= t_0$ 
with an amplitude of $\phi_0$. 
Then we have $I_0 \simeq m \phi_0^2$, 
which is much smaller than $m v^2 \simeq m \phi_0^2 (t /t_0)^{3p}$ for $t \gg t_0$. 
Therefore, 
what we need to calculate is the following integral: 
\beq
 \sqrt{I(\infty)} - \sqrt{I (t_0)}= - \int_{t_0}^\infty \dd t \sqrt{\frac{m}{2}} \dot{v} \sin \theta, 
\eeq
where $t_0$ is an arbitrary time much before $\tilde{m} (t) \simeq m_\phi (t)$. 
We need to specify a model to calculate this integral.

\subsection{Examples}

\subsubsection{Large Hubble-induced mass}

Here let us check that our method reproduces the result in Ref.~\cite{Linde:1996cx}, 
by taking $m_\phi = {\rm const.}$ and $\tilde{m} (t) = C H(t)$ with $C \gg 1$. 
This can be realized when 
we consider a modulus that has interaction with an inflaton 
with a cutoff scale below the Planck scale. 
In this case, we can take the limit of $t_0 \to 0$ 
because the axion initially stays at a certain VEV due to the Hubble friction effect 
(i.e., $I (t_0)=0$ for $t_0 \to 0$).

\begin{figure}[t]
\centering 
\includegraphics[width=.45\textwidth, bb=0 0 842 595
]{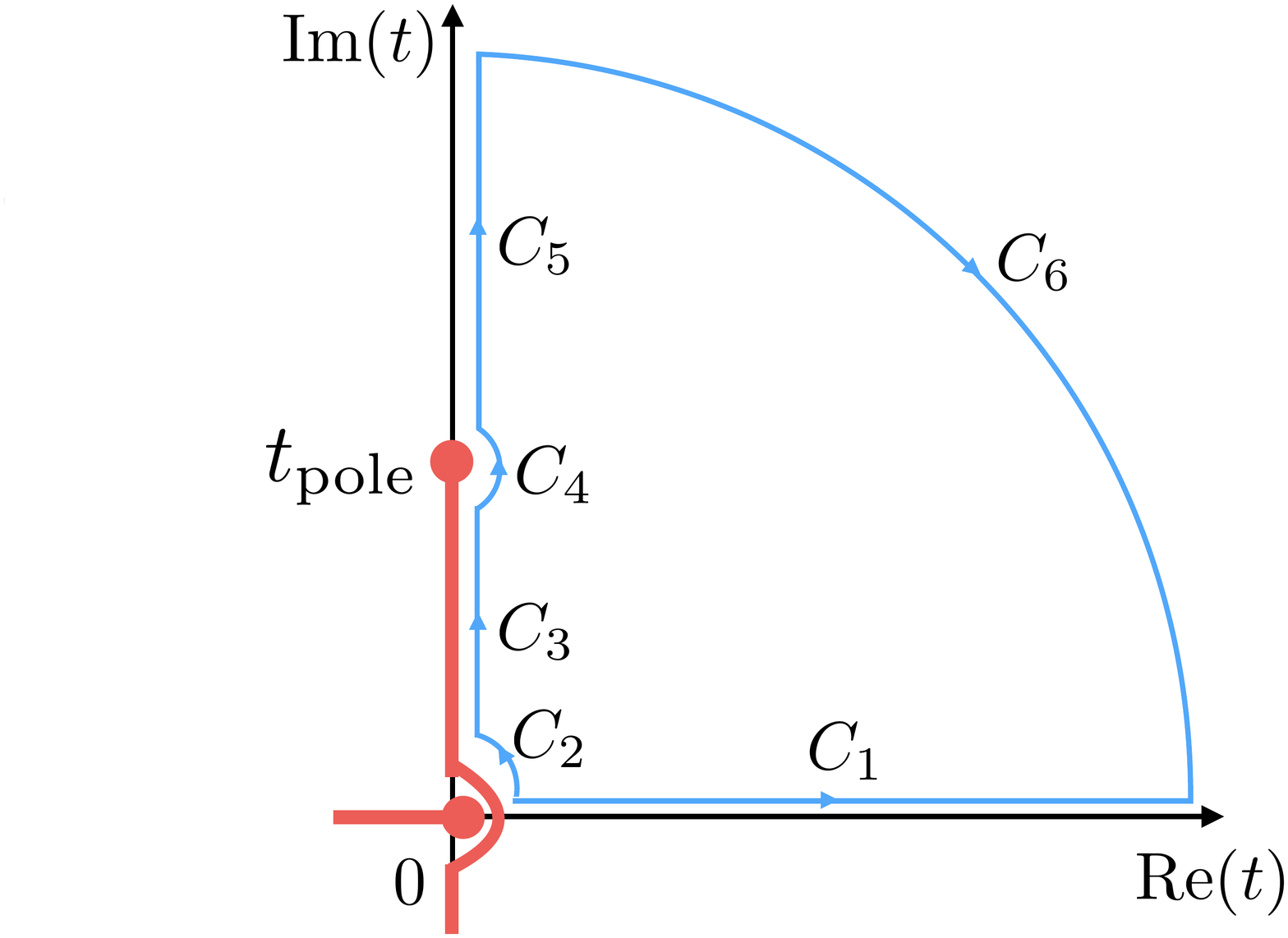} 
\hfill
\includegraphics[width=.45\textwidth, bb=0 0 842 595
]{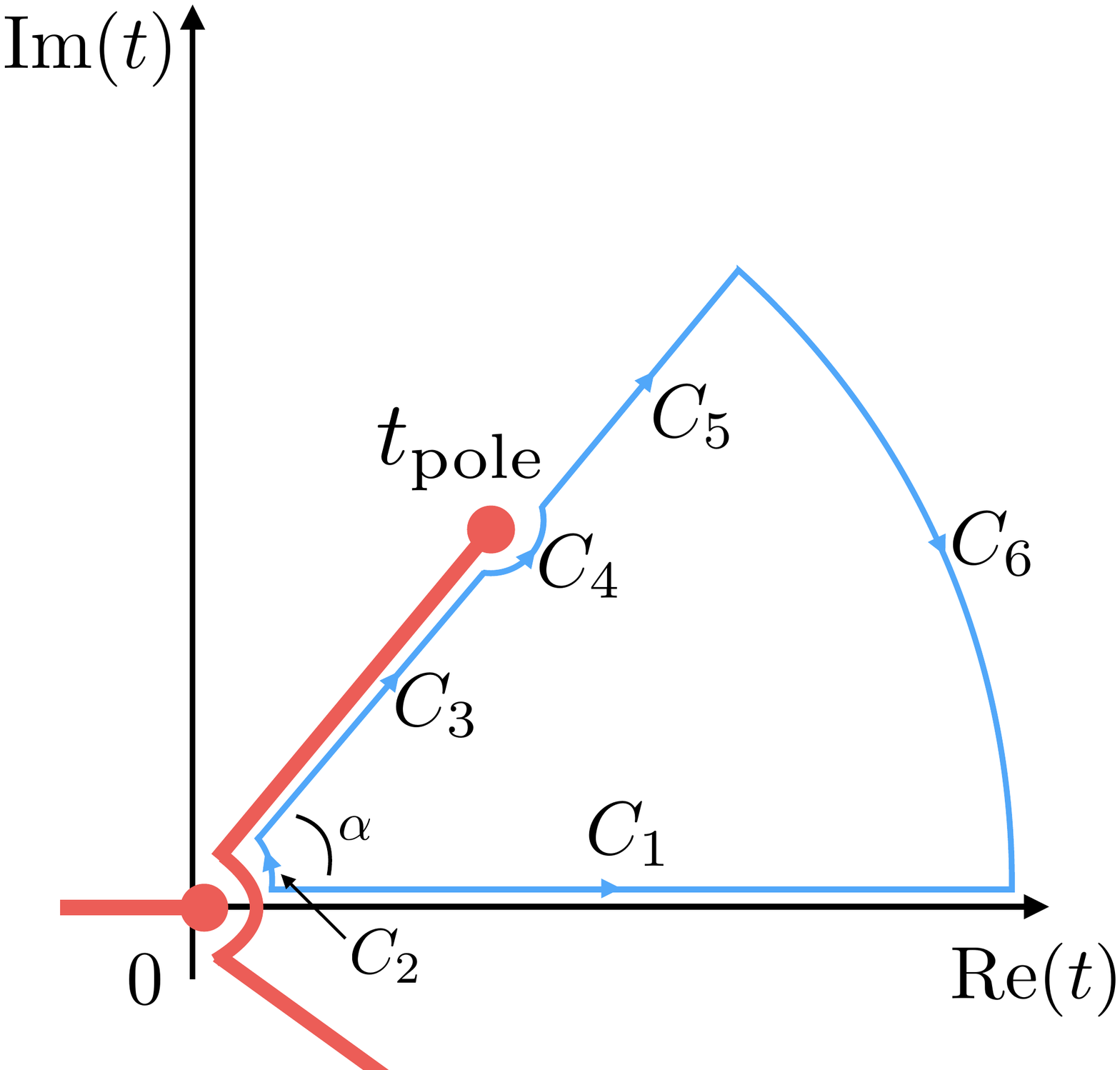} 
\caption{\small
Integration contours in the complex $t$ plane for the cases of $m_a = {\rm const.}$ (left panel) 
and $m_a \propto t^{n/2}$ (right panel). 
The red lines represent branch cuts 
and the red points are poles. 
  \label{fig:contour}
}
\end{figure}

We decompose the oscillation factor such as $\sin \theta = (e^{i \theta} - e^{-i \theta}) / 2 i$, 
and calculate each term separately. 
The integrand has poles and branch cuts in the complex $t$-plane, 
so that we replace the integration contour in the complex plane as shown in 
the left panel in Fig.~\ref{fig:contour}: 
\beq
 \frac{i}{2 \sqrt{2}} \int_{C_1} \dot{v} \sqrt{m}  e^{i \theta} \dd t 
 =  \frac{i}{2 \sqrt{2}} \int_{C_2 + C_3 + C_4 + C_5 + C_6} \dot{v} \sqrt{m} e^{i \theta} \dd t. 
 \label{contour}
\eeq
The integrand is given as 
\beq
 \dot{v} \sqrt{m} e^{i \theta} 
 = 
 \frac{V}{\tilde{C}^{3/2}} \frac{t^{3p/2 - 3/2}}{\lmk 1 + \frac{t^2}{z_{\rm pole}^2} \rmk^{7/4}} 
 \lkk \frac{3p}{2} + \lmk \frac{3p}{2} -2 \rmk \frac{t^2}{ z_{\rm pole}^2} \rkk e^{i \theta}, 
\eeq
where we define 
\beq
 &&z_{\rm pole} \equiv \tilde{C} / m_\phi
 \\
 &&\tilde{C} \equiv p \sqrt{3/2 (1/p - 3/2) + C^2 } \simeq p C 
 \\
 &&V \equiv p^2 C^2 \phi_0 t_0^{-3p/2}.
\eeq

The parameter $\theta$ is calculated from 
\beq
 \theta (t) = \int_{t_0}^t m(t') \dd t'. 
\eeq
On the imaginary axis of $t$, 
it is given by 
\beq
 \theta (t) 
 = 
 \left\{
\bea{ll}
 \int_{C_2} m(t') \dd t' 
 - \int_{z_0}^{z} \dd z' \sqrt{-m_\phi^2 + \tilde{C}^2/z'^2}
 ~~{\rm for}~~ z < z_{\rm pole} 
\vspace{0.5cm}\\
 \int_{C_2+C_3+C_4} m(t') \dd t' 
 + i \int_{z_{\rm pole}}^z \dd z' \sqrt{m_\phi^2 - \tilde{C}^2/z'^2}
 ~~{\rm for}~~ z > z_{\rm pole} 
\eea
\right.
\label{theta calc}
\eeq
where $t= i z$ and $t_{\rm pole} \equiv i z_{\rm pole}$ is the pole location in the complex $t$-plane 
and $z_0 \equiv t_0$. 
The integral for the contour $C_2$ can be calculated as 
\beq
 \int_{C_2} m(t') \dd t' 
 &=& \int_{C_2} \sqrt{m_\phi^2 t'^2 +\tilde{C}^2} \frac{\dd t'}{t'} 
 \\
 &=& \frac{\pi \tilde{C}}{2} i. 
 \label{eq:c2}
\eeq
The second terms in \eq{theta calc} are calculated as 
\beq
  - \int _{z_0}^{z} \dd z' \sqrt{-m_\phi^2 + \tilde{C}^2/z'^2}
  = - \tilde{C} \sqrt{1 - z^2 /z_{\rm pole}^2} 
  - \frac{\tilde{C}}{2} \ln \lkk \frac{1 - \sqrt{1 - z^2 / z_{\rm pole}^2}}{1 + 
  \sqrt{1 -z^2 / z_{\rm pole}^2}} \rkk + {\rm const.} 
  \nonumber\\ 
  \label{oscillation term}
  \\
   i \int_{z_{\rm pole}}^z \dd z' \sqrt{m_\phi^2 - \tilde{C}^2/z'^2}
   = i \tilde{C} \sqrt{z^2 /z_{\rm pole}^2 - 1} 
  + i \tilde{C} {\rm arctan} \lkk \frac{1}{\sqrt{z^2 / z_{\rm pole}^2}} \rkk 
  + {\rm const.}, 
  \nonumber\\ 
  \label{damping term}
\eeq
and $\int_{C_4} m(t') \dd t' = 0$ because $m(t) = 0$ at $t = t_{\rm pole}$.

Now we can calculate \eq{contour}. 
First, it is easy to see that the integral of $C_2$ in \eq{contour} is proportional to $t_0^{3p/2 - 1/2}$, 
so that it can be neglected in the limit of $t_0 \to 0$ for $p > 1/3$. 
Since we are interested in the radiation dominated era, 
where $p = 1/2$, this condition is satisfied.

Second, the integral of the contour $C_3$ 
has an oscillation term \eq{oscillation term} from $e^{i \theta}$. 
The time scale of this oscillation term ($\propto \tilde{C}$) is 
mush larger than that of the prefactor in the integrant $\dot{v} \sqrt{m}$, 
so that the integral of $C_3$ contour 
is suppressed by the oscillation term. 
However, when $t$ is sufficiently close to the pole ($\abs{t - i z_{\rm pole}} \lesssim z_{\rm pole}/\tilde{C}^{2/3}$), 
the oscillation term becomes constant, and the integral gives a large contribution. 

Finally, 
the integral of the contour $C_5$ and $C_6$ 
has a damping term \eq{damping term} from $e^{i \theta}$, 
so that 
the integral of $\int_{C_5 + C_6} e^{i \theta}$ 
is highly suppressed by the additional exponential factor. 
However, when $t$ is sufficiently close to the pole ($\abs{t - i z_{\rm pole}} \lesssim z_{\rm pole}/\tilde{C}^{2/3}$), 
the damping term becomes constant, and the integral gives a large contribution.

Therefore, the integral at the vicinity of the pole 
gives the main contribution of Eq.~(\ref{contour}). 
We should calculate 
\beq
 &&\int_{{\abs{t - i z_{\rm pole}}< z_{\rm pole} / \tilde{C}^{2/3}}} \dot{v} \sqrt{m} e^{i \theta} \dd t 
 \\
 &&\simeq 
 \frac{2V}{\tilde{C}^{3/2}} e^{i \theta( t = i z_{\rm pole})} 
 \lmk i z_{\rm pole} \rmk^{3p/2 - 3/2} 
 \int_{{\abs{t - i z_{\rm pole}}< z_{\rm pole} / \tilde{C}^{2/3}}} \lmk 1 + \frac{t^2}{z_{\rm pole}^2} 
 \rmk^{-7/4} \dd t. 
 \nonumber
 \\
\eeq
where the integral should be taken on the contour $C_3 + C_4 + C_5$. 
The integral can be calculated as 
\beq
 &&\int_{{\abs{t - i z_{\rm pole}}< z_{\rm pole} / \tilde{C}^{2/3}}} \lmk 1 + \frac{t^2}{z_{\rm pole}^2} 
 \rmk^{-7/4} \dd t
 \nonumber\\
&&=
 \frac{4}{3} \lmk \frac{z_{\rm pole}}{2} \rmk^{7/4} \lmk \frac{\tilde{C}^{2/3}}{z_{\rm pole}} \rmk^{3/4} 
 e^{-7/8 \pi i} \lmk e^{3/8 \pi i} - e^{-3/8 \pi i} \rmk. 
\eeq

We can calculate $\int \dot{v} \sqrt{m} e^{-i \theta} \dd t$ 
in the same way. 
Combining these results, 
we obtain 
the adiabatic invariant such as 
\beq
 I \simeq 
 \frac{\tilde{C}^{3p+1}}{m_\phi^{3p - 1}} \phi_0^2 t_0^{-3p} e^{- \pi p C}
 \label{I result}
\eeq
where we assume $I (t_0) = 0$ 
and omit $\mathcal{O}(1)$ factors. 
Using $I \simeq m_\phi \cphi^2$ for $C H \ll m_\phi$, 
we obtain 
\beq
 \phi \equiv \lmk \frac{t_0}{t} \rmk^{3p/2} \cphi 
 \simeq 
 p^{3/2} C^{(3p+1)/2} 
 \phi_0 
 \lmk \frac{p}{t m_\phi} \rmk^{3p/2} e^{-\pi p C/2}, 
\eeq
where we use $\tilde{C} \simeq p C$. 
All parameter dependences are consistent with the one derived by Linde. 
Thus we can solve the moduli problem when it has an effective mass 
term $C H(t)$ with $C \sim 30$~\cite{Linde:1996cx}.

To sum up, 
the exponential factor comes from \eq{eq:c2} 
and the other factors should be calculated around the vicinity of 
the pole $t = i z_{\rm pole}$. 
These facts allow us to easily estimate the resulting adiabatic invariant.

Finally, we rewrite the result in terms of 
the number density to entropy density ratio for later convenience: 
\beq 
 \frac{n_\phi}{s} 
 &\equiv& \left. \frac{m_\phi \phi^2/2}{s} \right\vert_{\rm t \gg z_{\rm pole}}
 \\
 &=& \frac{45 }{2 \pi^2 g_{*s}}
 \lmk \frac{4 \pi^2 g_*}{90 M_{\rm pl}^2} \rmk^{3/4} 
 \times t_0^{3/2} I , 
\eeq
where we use $p = 1/2$, 
and $g_{*s}$ and $g_*$ are 
effective relativistic degrees of freedom for entropy and energy density, respectively.

\subsubsection{Application to an axion model}

Now we calculate the abundance in the case where $m_\phi$ depends 
on time such as the case of QCD axion, where $m_\phi (t) \equiv A t^{pn/2}$ [see \eq{axion mass at T}]. 
In this subsection, we assume $\tilde{m}(t) = C H(t)$, which is considered in Refs.~\cite{Folkerts:2013tua, Takahashi:2015waa}. 
This can be realized when 
we introduce a term like 
\beq
 \mathcal{L} \supset c_R^2 R M_{\rm pl}^2 \cos \lmk \frac{a}{f_a} - \theta_R \rmk, 
\eeq
where $R$ is the Ricci scalar. 
In the radiation dominated era, this term gives $\tilde{m}(t) = C H(t)$ 
with $C = 1.6 c_R \alpha_s M_{\rm pl} / f_a$, where $\alpha_s$ is the fine-structure constant of 
SU(3)$_c$ gauge interaction.

As shown in the right panel in Fig.~\ref{fig:contour}, 
the pole location is given by $t = e^{i \alpha} z_{\rm pole}$, 
where 
\beq 
 &&\alpha = \pi / (2 + p n)
 \\
 &&z_{\rm pole} = \lmk \frac{\tilde{C}}{A} \rmk^{2/(2+pn)}
 \\
 &&\tilde{C} \equiv p \sqrt{3/2 (1/p - 3/2) + C^2 } \simeq p C. 
\eeq
In this case, 
$i \theta (t = e^{i \alpha}  z_{\rm pole})$ is calculated as 
\beq
 i \theta (t = e^{i \alpha}  z_{\rm pole}) = - \alpha p C + ({\rm imaginary \ part}). 
\eeq

The integral of \eq{contour} is again dominated by the contribution in the vicinity of the pole: 
$\abs {t - e^{i \alpha} z_{\rm pole}} < z_{\rm pole} / ( \tilde{C} \sqrt{2 + pn})^{2/3}$. 
Then we estimate the adiabatic invariant such as 
\beq
 I &\simeq& 
 \frac{\tilde{C}^{2} z_{\rm pole}^{3p -1}}{pn + 2} 
 t_0^{-3p} \phi_0^2 
 e^{- 2 \alpha p C} 
 \\
 &\simeq& 
 \frac{\tilde{C}^{2 (pn+1+3p)/(2+ pn)}}{pn + 2} A^{2(1-3p)/(2+pn)} t_0^{-3p} \phi_0^2 
 e^{- 2 \alpha p C}. 
\eeq
Therefore, 
an effective mass of $\tilde{m}(t) = C H(t)$ 
suppresses the axion abundance efficiently when $C \gg 1$.

\subsubsection{Application to a generic model}

Finally, 
we calculate the abundance in the case of $m_\phi (t) = A t^{pn/2}$ 
and $\tilde{m} (t) = C' t^{- d}$, 
where $d$ is a constant parameter. 
The following calculation should reproduce the previous result 
if we take $d = 1$ and $C' = C p$.

In this case, $m^2 (t)$ is given by 
\beq
 m^2 (t) = A^2 t^{pn} + \frac{3}{2} \lmk \frac{1}{p} - \frac{3}{2} \rmk H^2(t) 
 + 
 C'^2 t^{-2 d}. 
\eeq
We neglect the $H^2 (t)$ term because we are interested in 
the change of adiabatic invariant 
at a time around $A^2 t^{pn} \sim C'^2 t^{-2d} \gg H^2 (t)$. 
Note that $t_0$ cannot be taken to be $0$ 
because $C'^2 t^{-2d}$ is larger than $H^2(t)$ only after a certain time. 
Thus we take, say, $t_0 \approx z_{\rm pole} / 10$, 
though our result is independent of this value for $t_0 \ll z_{\rm pole}$. 
Here, $z_{\rm pole}$ is given by 
\beq
 z_{\rm pole} \equiv \lmk \frac{C'}{A} \rmk^{2/(2 d + pn)}. 
\eeq
The pole location is given by $ t = e^{i \alpha} z_{\rm pole}$, 
where 
\beq
 \alpha = \frac{\pi}{ 2 d + p n}.
\eeq

First, let us calculate the imaginary part of $\theta (t_{\rm pole})$. 
The integral on the contours $C_2$ and $C_3$ is calculated as 
\beq
 {\rm Im} [\theta (t_{\rm pole}) ]
 &=& 
 {\rm Im} \int_{C_2+C_3} m(t') \dd t' 
 \nonumber\\
 &\simeq& 
 C' z_{\rm pole}^{1-d} 
 \,
 {\rm Im} \lkk 
 \int_{t_0/z_{\rm pole}}^{t_0/z_{\rm pole} e^{i \alpha}} \dd t' t'^{-d} 
 + e^{i (1-d) \alpha} \int_{t_0/z_{\rm pole}}^1 \dd t' \sqrt{t'^{ - 2 d} - t'^{pn}} 
 \rkk
 \nonumber\\
 &=& 
 \alpha C' z_{\rm pole}^{1-d} D \frac{\sin (1-d)\alpha}{(1-d) \alpha}, 
\eeq
where in the second line we define 
\beq
 D \equiv 
  \lkk 
 \lmk \frac{t_0}{z_{\rm pole}} \rmk^{1-d} + 
 (1-d) \int_{t_0/z_{\rm pole}}^1 \dd t' \sqrt{t'^{ - 2 d} - t'^{ pn}} 
 \rkk. 
 \label{D}
\eeq
We can evaluate $D$ numerically for given parameters and 
we find that it is almost independent of $t_0 / z_{\rm pole}$ 
for $t_0 / z_{\rm pole} \ll 1$ 
and is about unity for $(2d + p n ) \gg 1$.

The integral of \eq{contour} is again dominated by the contribution in the vicinity of 
the pole: 
\beq
 \abs{t - e^{i \alpha} z_{\rm pole} }< z_{\rm pole} \lmk \frac{z_{\rm pole}^{d-1}}{ C' \sqrt{2 d + pn} }\rmk^{2/3}. 
\eeq 
Then we estimate the adiabatic invariant such as 
\beq
 I \simeq 
 \frac{C'^2 }{2 d + pn } 
 z_{\rm pole}^{3p - 2d +1} 
 t_0^{-3p} \phi_0^2  
 e^{- 2 {\rm Im} [\theta (t_{\rm pole})]}. 
 \label{I result}
\eeq

It is convenient to rewrite this result as 
the number density to entropy density ratio: 
\beq 
 \frac{n_\phi}{s} 
 &\equiv& \left. \frac{m_\phi \phi^2/2}{s} \right\vert_{\rm t \gg z_{\rm pole}}
 \\
 &=& \frac{45 }{2 \pi^2 g_{*s}}
 \lmk \frac{4 \pi^2 g_*}{90 M_{\rm pl}^2} \rmk^{3/4} 
 \frac{2 C'^2}{n+4d} 
 z_{\rm pole}^{5/2-2d} 
 \phi_0^2
  e^{- 2 {\rm Im} [\theta (t_{\rm pole})]}
  \label{Y_phi}
\eeq
where we use $p = 1/2$, 
and $g_{*s}$ and $g_*$ are 
effective relativistic degrees of freedom for entropy and energy density, respectively. 
Note that ${\rm Im} [\theta (t_{\rm pole})] \propto C' z_{\rm pole}^{1 - d} \simeq \tilde{m} / H(t = z_{\rm pole}) \gg 1$.

\section{QCD axion and adiabatic suppression mechanism
\label{sec:QCD axion}}

The QCD axion potential is protected by the PQ symmetry,
and therefore, it is difficult to give the axion 
a time-dependent effective mass term, which is needed to realize the adiabatic suppression mechanism. 
In this section, we consider the cosmological history of 
the QCD axion or string axion that acquires an effective mass due to 
the Witten effect, 
using the result given in the previous section.

\subsection{Axion dynamics with monopole DM 
\label{sec:dynamics}}

Here we consider a cosmological history of the axion in the presence of hidden monopoles of U(1)$_H$. 
We assume that the PQ symmetry is anomalous under the hidden U(1)$_H$, 
which then implies that the axion acquires an effective mass due to the Witten effect (see App.~\ref{sec:witten}). 
In this section, we consider the case where 
U(1)$_H$ is not broken, and monopole is stable and can be DM~\cite{Murayama:2009nj}.%
\footnote{
A scenario where t'Hooft-Polyakov monopole accounts for DM was discussed in Ref.~\cite{Baek:2013dwa, Khoze:2014woa}, 
where massive gauge bosons are another component of DM. 
}

Suppose that monopoles are produced at a temperature of $T_m$ ($T_m \gg \LQCD$) 
with an initial number density of $n_M (T_m)$. 
Its number density decreases with time due to the cosmic expansion, so that 
the ratio to the entropy density is constant: 
\beq
 Y_M \equiv \frac{n_M}{s} = \const.
\eeq

In the presence of monopoles, 
the Witten effect gives an effective mass of the axion [see \eq{FP effect}]~\cite{Witten:1979ey, Fischler:1983sc}: 
\beq
 \mam (T) = 2 \beta \frac{n_M (T)}{f_{a,H}}, 
\eeq
where $\beta$ is a constant defined by \eq{beta-wo cutoff} or \eq{beta_cutoff}. 
Here, the axion decay constant associated with U(1)$_H$ is denoted as $f_{a,H}$, 
which may be different from $f_a$ by a rational number 
because of the difference of periodicity of the axion (or so-called the domain wall number).%
\footnote{
$f_{a,H}$ can be enhanced by many orders of magnitude 
with respect to $f_a$ in the clockwork mechanism~\cite{Higaki:2015jag, Higaki:2016yqk}. 
}
Specifically, they are related to each other by $f_{a,H} = (N_{\rm DW} / N_{a,H}) f_a$, 
where $N_{\rm DW}$ and $N_{a,H}$ are the domain wall numbers of the axion 
in terms of the QCD instanton effect and the Witten effect, respectively. 
This implies that the ratio 
between the effective axion mass and the Hubble parameter 
increases with time such as 
\beq
 \frac{\mam (T)}{H^2 (T)} = 2 \beta Y_M \frac{s}{H^2 f_{a,H}}, 
 \label{fraction}
\eeq
which is approximately proportional to $T^{-1}$ in the radiation dominated era. 
The ratio becomes unity at a temperature of $T = T_{\rm osc, 1}$ and time $t = t_{\rm osc,1}$, 
which is given by 
\beq
T_{\rm osc, 1} 
 & \simeq& 
 Y_M 
 \frac{8 \beta \Mpl^2}{f_{a,H}} \\
 & \simeq& 2 \times 10^9 {\rm\,GeV} \beta \lrfp{m_M}{1 \TeV}{-1} 
 \lrf{\Omega_M h^2}{0.12} \lrfp{f_{a, H}}{10^{16}\,{\rm GeV}}{-1}, 
\eeq
where $\Omega_M$ is the relic density of monopole: 
\beq
 \Omega_M h^2 \simeq \frac{m_M Y_M}{3.6 \eV}. 
 \label{OmegaM}
\eeq

In the case of the t'Hooft-Polyakov monopole, 
$\beta$ is given by \eq{beta_cutoff} ($\beta = \alpha_H/(32 \pi^2 r_c f_{a,H})$), 
where $\alpha_H$ is the fine-structure constant of U(1)$_H$ 
and $r_c$ is an electric screening scale of an electrically charged particle.%
\footnote{
The monopole may be a fundamental particle as considered in Refs.~\cite{Yamada:2015waa,Yamada:2016jgg,Kamada:2016ois}. 
In this case, we should use \eq{beta-wo cutoff} instead of \eq{beta_cutoff}, 
which leads to a larger Witten effect. 
} 
We may take 
$r_c = m_W^{-1} = 1/ (e v) \simeq 1/ \alpha_H m_M$, 
where $m_W$ is the mass of SU(2) gauge bosons, 
$e$ is the hidden gauge coupling constant, 
and $v$ is the SU(2) breaking scale. 
Then we obtain 
\beq
T_{\rm osc, 1} \simeq 
65 {\rm\,GeV} \alpha_H^2 \lrf{\Omega_M h^2}{0.12} \lrfp{f_{a, H}}{10^{12}\,{\rm GeV}}{-2}. 
\label{T_osc1}
\eeq
In this paper, we require that the ratio becomes unity before the QCD phase transition, 
i.e., $T_{\rm osc, 1} \gtrsim T_{\rm osc, 0}$, 
which condition is rewritten as 
\beq
f_a \lesssim 9 \times 10^{12}\,{\rm GeV} \alpha_H^{1.1} 
\lrfp{N_{a, H}}{N_{\rm DW}}{1.1}
\lrfp{\Omega_M h^2}{0.12}{0.55}, 
\label{condition1}
\eeq
for the case of the t'Hooft-Polyakov monopole. 
Note that $f_a$ may be different from $f_{a,H}$ by a rational factor.

The Witten effect should be sufficiently small so that 
the axion VEV cancels the undesired strong CP phase at present. 
We can check it as follows: 
\beq
 \left. \frac{\mam}{m_a^2} \right\vert_{t_p} \simeq 
 10^{-29} \beta 
 \lrf{\Omega_M h^2}{0.12} 
 \lrfp{m_M}{1 \TeV}{-1} 
 \lrf{f_a}{10^{16} \GeV}, 
\eeq
where the left hand side is evaluated at the present time $t_p$. 
Note that $\beta \ll 1$. 
Since this fraction is much smaller than $10^{-10}$ at present, 
the strong CP problem is solved even in the presence of monopole.

\subsection{Axion abundance}

We consider a scenario where the PQ symmetry is broken before inflation. 
In this case, the axion stays at a certain phase during and after inflation. 
After inflation ends and monopoles are produced, 
the ratio of \EQ{fraction} increases and reaches about unity at $T = T_{\rm osc,1}$. 
Then the axion starts to oscillate around the minimum (i.e., $\la a \ra = \theta$), 
which is determined by the hidden sector.

The axion starts to oscillate by the Witten effect 
before the QCD phase transition when \eq{condition1} is satisfied. 
The resulting axion energy density is given by $H^2 (T_{\rm osc, 1}) \theta_{\rm ini, H}^2 f_{a, H}^2/2$ at the onset of oscillation, 
where $(\theta_{\rm ini, H} + \theta)$ is the initial misalignment angle of axion. 
This gives the initial value of adiabatic invariant: 
\beq
 I_0 = \frac{1}{2} H (T_{\rm osc, 1}) \theta_{{\rm ini}, H}^2 f_{a, H}^2 \lmk \frac{t_{\rm osc,1}}{ t_0} \rmk^{3/2}, 
\eeq
which represents the comoving number density of axion normalized at $t = t_0$. 
Note that we include the factor of $\lmk t_{\rm osc,1} / t_0 \rmk^{3/2}$ 
because we normalize the adiabatic invariant at $t = t_0$.

Although the effective mass of the axion \EQ{FP effect} decreases with time, 
its comoving number density, which is an adiabatic invariant, is approximately conserved after the onset of oscillation. 
In particular, 
the effect of the QCD instanton around the time of QCD phase transition 
does not affect the number density of axion so much. 
The resulting amount of induced axion at the QCD phase transition is 
given by \eq{I result} with $d = 3p/2$: 
\beq
 \Delta I \simeq 
 \frac{2C'^2}{n+3} 
 z_{\rm pole}
 t_0^{-3/2} \phi_0^2 
 e^{- 2 \alpha C' z_{\rm pole}^{1/4} D}. 
\eeq
where $D$ ($\simeq 1$) is given by \eq{D} and 
\beq
 &&\phi_0 = \theta f_{a, H}
 \\
 &&C' = H (t_{\rm osc,1}) t_{\rm osc,1}^{3/4}
 \\
 &&z_{\rm pole} = \lmk \frac{H^2 (t_{\rm osc,1})}{m_a^2 (t_{\rm osc,1})} 
 \rmk^{2/(n+3)} t_{\rm osc,1}
 \\
 &&\alpha = \frac{2\pi}{n+3}. 
\eeq
Here and hereafter we use $p =1/2$. 
The induced $\Delta I$ is smaller than $I_0$ when 
\beq
  \frac{2 C'^2 z_{\rm pole}^{1/2}}{n+3} 
 \lmk \frac{z_{\rm pole}}{t_{\rm osc, 1}} \rmk^{1/2} 
 e^{- 2 \alpha C' z_{\rm pole}^{1/4}} 
 \lesssim 1
 \\
 \leftrightarrow
  \frac{H^2 (t_{\rm osc,1})}{m_a^2 (t_{\rm osc,1})} 
  \gtrsim 
  \lkk \frac{1}{\pi} \ln \lmk  \frac{H^2 (t_{\rm osc,1})}{m_a^2 (t_{\rm osc,1})}  \rmk 
  \rkk^{2(n+3)}, 
\eeq
where we use $\alpha = 2\pi/(n+3)$ 
and assume $\theta_{{\rm ini},H} \sim \theta$ ($= \mathcal{O}(1)$). 
This inequality 
is usually satisfied. 
However, we should note that 
we assume $H^2 (t_{\rm osc,1}) \gg m_a^2 (t_{\rm osc,1})$ 
in the above calculation. 
Therefore, 
we conclude that 
the resulting axion abundance is determined by $I_0$
when $H (t_{\rm osc,1}) \gtrsim m_a (t_{\rm osc,1})$.

When the adiabatic suppression mechanism works, 
the present axion number density is determined by $I_0$ and is given as 
\beq
 \frac{n_a}{s} 
 &\simeq& 
 \frac{ H_{\rm osc, 1} \theta_{\rm ini, H}^2 f_{a, H}^2/2}{s (T_{\rm osc, 1})} 
 \\
 &\simeq& 
 \sqrt{\frac{45}{32 \pi^2 g_*}} \frac{\theta_{\rm ini, H}^2 f_{a, H}^2}{T_{\rm osc, 1} \Mpl}. 
 \label{Y_a}
\eeq
The axion density parameter is thus given by 
\begin{align}
 \Omega_a h^2 
& \simeq 
 2 \times 10^{-14} \theta_{\rm ini, H}^2 \frac{f_{a, H}}{T_{\rm osc, 1}} 
 \lmk \frac{N_{\rm DW}}{N_{a,H}} \rmk 
 \\
& \simeq 
3 \times 10^{-4}\, \frac{\theta_{\rm ini, H}^2}{\alpha_H^{2}} 
\lmk \frac{N_{\rm DW}}{N_{a,H}} \rmk 
\lrf{0.12}{\Omega_M h^2} \lrfp{f_{a, H}}{\GEV{12}}{3}. 
\label{Omegaa2}
\end{align}
Since the total abundance should be smaller than the observed DM abundance, 
we obtain an upper bound on the axion decay constant: 
\beq
 f_{a, H} \lesssim 4 \times 10^{12} \GeV 
 \lmk \frac{\alpha_H}{\theta_{\rm ini, H}} \rmk^{2/3} 
 \lmk \frac{N_{a,H}}{N_{\rm DW}} \rmk^{1/3}, 
 \label{fa bound}
\eeq
where we used $(\Omega_a h^2) (\Omega_M h^2) \le (\Omega_a h^2 + \Omega_M h^2)^2/4 \lesssim 0.12^2 / 4$.

\begin{figure}[t]
\centering 
\includegraphics[width=.60\textwidth, bb=0 0 360 343]{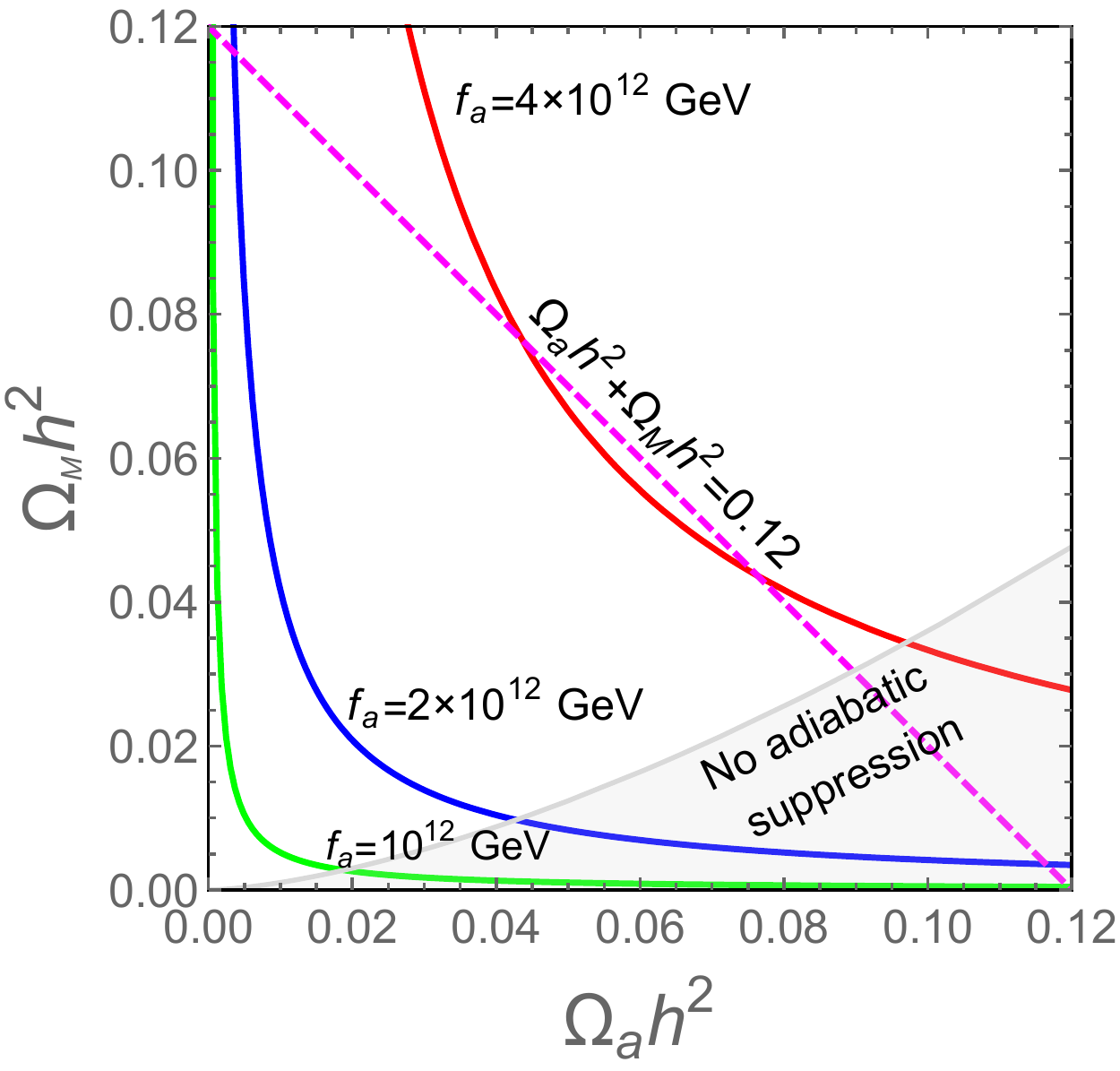} 
\caption{
Relation between the axion and monopole abundances.
We take $\fa =1 \times 10^{12} \GeV^3$ (green curve), 
$2 \times 10^{12} \GeV$ (blue curve), 
and $4 \times 10^{12} \GeV$ (red curve). 
We assume $\theta_{\rm ini, H} = 1$, $\alpha_H = 1$, and $N_{\rm DW} = N_{a,H}$. 
The diagonal (magenta) dashed line represents the observed DM abundance. 
In the light-gray shaded region, 
the adiabatic suppression does not work 
and the results should not be trusted. 
}
  \label{fig1}
\end{figure}

We show contours of the axion and monopole abundance in Fig.~\ref{fig1}. 
We take $\fa =1 \times 10^{12} \GeV$ (green curve), 
$2 \times 10^{12} \GeV$ (blue curve), 
and $4 \times 10^{12} \GeV$ (red curve). 
We assume $\theta_{\rm ini, H} = 1$, $\alpha_H = 1$, and $N_{\rm DW} = N_{a,H}$. 
The observed  DM abundance can be explained at the intersection point 
between the contour curve and the diagonal magenta dashed line 
for each value of $f_a$. 
Note that the adiabatic suppression mechanism works 
for the case of $T_{\rm osc,1} \gtrsim T_{\rm osc, 0}$, 
so that we should restrict ourselves to the case of 
\beq
 \Omega_a h^2 \lesssim 0.22 \ \theta_{\rm ini, H}^2 \alpha_H^{1.3} 
 \lmk \frac{N_{a, H}}{N_{\rm DW}} \rmk^{-0.69} 
 \lmk \frac{\Omega_M h^2}{0.12} \rmk^{0.65}. 
 \label{adiabatic condition2}
\eeq
This is shown in Fig.~\ref{fig1} as the unshaded region. 
The adiabatic suppression mechanism does not work 
in the light-gray shaded region, 
in which case 
the coherent oscillation of axion is induced at the QCD phase transition 
and the resulting axion abundance is given by the sum of Eqs.~(\ref{Omegaa}) and (\ref{Omegaa2}).

Finally, we comment on the excited states of the monopole (i.e., Julia-Zee dyons). 
At the monopole production process, 
some excited states of a monopole with nonzero hidden-electric charges 
may be produced~\cite{Julia:1975ff}. 
This is considered to be the case especially if the axion field takes different values 
over the space, e.g., around the axionic cosmic string. 
In the scenario considered in this subsection, however, 
the effective theta parameter of the hidden U(1)$_H$ (i.e., axion VEV) is constant 
in the whole Universe at the monopole production process, 
so that we expect that 
only monopoles are produced at that time. 
Thus the axion acquires the space-independent quadratic mass term.

\subsection{Isocurvature problem}

When the PQ symmetry is broken before inflation, 
the axion acquires quantum fluctuations during inflation~\cite{Axenides:1983hj}. 
This is quantitatively written in terms of fluctuations of initial misalignment angle 
$\delta \theta_{\rm ini, H}$ such as 
\beq
 \delta \theta_{\rm ini, H} \approx \frac{H_{\rm inf}}{2 \pi f_{a, H}}, 
\eeq
where $H_{\rm inf}$ is the Hubble parameter during inflation. 
This results in isocurvature modes in density perturbations whose amplitude is given by 
\beq
 P_{\rm iso} \simeq \lmk \frac{\Omega_a }{\Omega_{\rm DM} } \rmk^2 
 \lmk \frac{H_{\rm inf}}{ \pi f_{a,H} \theta_{{\rm ini},H}} \rmk^2. 
 \label{Piso}
\eeq

Since the CMB temperature anisotropies are predominantly adiabatic, 
the Planck observation puts the constraint on the amplitude of isocurvature 
perturbations: 
\beq
 \frac{P_{\rm iso}}{P_{\rm ad}} \lesssim 0.037, 
\eeq
where the amplitude of adiabatic perturbations is measured as $P_{\rm ad} \simeq 2.2 \times 10^{-9}$~\cite{Ade:2015lrj}. 
As a result, 
the axion decay constant should satisfy 
\beq
 f_{a, H} 
 \gtrsim 3.4 \times 10^{4} H_{\rm inf} \frac{\Omega_a}{\Omega_{\rm DM}} \frac{1}{\theta_{\rm ini, H}},  
 \label{isocurvature constraint2}
\eeq
which can be rewritten by Eq.~(\ref{Omegaa2}) such as 
\beq
 H_{\rm inf} \lesssim 1.2 \times 10^{10} \GeV \alpha_H^2 \theta_{\rm ini, H}^{-1} 
 \lmk \frac{N_{a,H}}{N_{\rm DW}} \rmk 
 \lrf{\Omega_M h^2}{0.12} \lrfp{f_{a, H}}{\GEV{12}}{-2}. 
\eeq
One can see that the isocurvature constraint on the inflation scale is greatly relaxed for given $f_{a}$
and the initial misalignment~\cite{Kawasaki:2015lpf}. For instance, in the ordinary scenario without
the Witten effect, the upper bound on $H_{\rm inf}$ is of order $10^7$\,GeV for $f_a = 10^{12}$\,GeV
and $\theta = O(1)$. This is because the axion abundance is suppressed 
by the Witten effect and it can be smaller than the observed DM abundance without fine-tuning the initial
misalignment angle. The hidden monopoles are a prime candidate for the major DM component. 
Note that,  in this scenario,  the perturbation of axion energy density, 
$\delta \Omega_a/\Omega_a$, is not suppressed, and therefore, this effect cannot be mimicked by making
the initial displacement smaller since it would enhance $\delta \Omega_a/\Omega_a$.

\subsection{Constraints on monopole DM} 

When the hidden U(1)$_H$ is unbroken, 
the monopole can be DM~\cite{Murayama:2009nj, Baek:2013dwa}.%
\footnote{
When we consider t'Hooft-Polyakov monopole originated from the SSB of SU(2) to U(1)$_H$, 
there are massive gauge bosons as well as monopoles. 
This theory has been investigated in Ref.~\cite{Baek:2013dwa, Khoze:2014woa}, 
where they found that 
the massive gauge bosons are a dominant component of DM. 
}
Since monopoles interact with hidden photons, 
they are interacting massive DM. 
The cross section is calculated as~\cite{Khoze:2014woa}
\beq
 \sigma_T 
 &=& \frac{16 \pi \alpha_M^2}{ m_M^2 v^4} 
 \log \lmk 1 + \frac{m_M^4 v^4}{8 \pi  \alpha_M^2 \rho_M} \rmk 
 \label{sigma_T} 
 \\ 
 &\simeq& 
 0.4 \cm^2 /{\rm g} \ m_M \alpha_M^{2} 
 \lrfp{v}{10 \km / {\rm s}}{-4} 
 \lrfp{m_M}{1 \PeV}{-3} 
\eeq
for monopole DM, where $\alpha_M \equiv g^2 / (4 \pi)$ ($ = 1/\alpha_H$) is the fine-structure constant 
for the magnetic charge 
and we take the log factor as $\approx 40$ in the second line.

There are several constraints on the self-scattering cross section 
depending on the DM velocity $v$. 
The Bullet cluster gives the upper bound such as $\sigma_T / m_M < 1.25 \cm^2 /{\rm g}$ 
for $v \sim 1000 \km / {\rm s}$~\cite{Clowe:2003tk, Markevitch:2003at, Randall:2007ph}. 
The most stringent constraint comes from the ellipticity of DM halo, 
where observations indicate that DM halos are somewhat elliptical. 
Since the self-interaction of DM results in a spherical DM halo, 
the cross section is restricted above by this consideration such as 
$\sigma_T / m_M \lesssim 0.1 -1 \cm^2 /{\rm g}$ 
for $v \sim 200 \km / {\rm s}$~\cite{Rocha:2012jg, Peter:2012jh}. 

The self-scattering of DM is well motivated 
to solve astrophysical problems: core-cusp problem 
and too-big-to-fail problem~\cite{Spergel:1999mh}. 
These can be solved when 
$\sigma_T / m_M < 0.1 - 10 \cm^2 /{\rm g}$ 
for $v \sim 10-30 \km / {\rm s}$~\cite{Zavala:2012us}. 
Also, 
the recent observation of Abel 3827 indicates 
$\sigma_T / m_M = 1.5 \cm^2 /{\rm g}$, 
which also motivates us to consider the self-interacting monopole DM~\cite{Kahlhoefer:2015vua} 
(see also Ref.~\cite{Massey:2015dkw}). 
These can be addressed when the mass of monopole is of order PeV scale 
for $\alpha_H =1$.

\section{
String axion with monopole annihilation
\label{sec:string axion}}

In this section, we consider the case that 
the monopoles disappear at a temperature of $T_{\rm ann}$ after 
the QCD phase transition but before the BBN epoch 
(i.e., $1 \MeV \lesssim T_{\rm ann} < \LQCD $). 
This can be realized by the SSB of U(1)$_H$ at the temperature of 
$T_{\rm ann}$ 
because each monopole and anti-monopole pair is attached by a cosmic string 
associated with the SSB of U(1)$_H$ 
and annihilate with each other~\cite{Nambu:1974zg}. 
In this case, 
the Witten effect turns off and the axion becomes massless at $T_{\rm ann}$. 
Since there is no monopole in the present Universe,%
\footnote{
We assume that the massive hidden photon decays into the SM particles
through a kinetic mixing between the hidden U(1)$_H$ and
hypercharge soon after the monopole annihilation. 
}
we are interested in the case that the axion is the dominant component of DM. 

When we require that the monopoles do not dominate the Universe before 
they disappear, 
we should satisfy 
\beq
 \left. \frac{\rho_M}{\rho_{\rm tot}} \right\vert_{T = T_{\rm ann}} \lesssim 1 
\eeq
where 
\beq
 \left. \frac{\rho_M}{\rho_{\rm tot}} \right\vert_{T = T_{\rm ann}} 
 &\simeq& 
 \lmk \frac{\Omega_M h^2}{2 \times 10^{7} } \rmk 
 \lmk \frac{T_{\rm ann}}{100 \MeV} \rmk^{-1}, 
\eeq
where we use $g_* \approx g_{s*}$ and define $\Omega_M h^2$ via the relation of 
\eq{OmegaM}. 
Or, we should satisfy 
\beq
 \lmk \frac{T_{\rm osc, 1}}{T_{\rm ann}} \rmk 
 \lesssim 1.2 \times 10^{4} 
 \alpha_H^{2} \lmk \frac{f_a}{10^{16} \GeV} \rmk^{-2}. 
 \label{condition for domination}
\eeq
where we assume t'Hooft-Polyakov monopole and 
use $\beta = \left. \beta \right\vert_{\rm with \  cutoff}$ 
with $r_c = 1/(\alpha_H m_M)$ [see \eq{beta_cutoff}].

On the other hand, we can also consider the case that 
the monopoles dominate the Universe 
before they annihilate. 
In this case, 
the annihilation of monopole generates entropy and dilute SM plasma. 
We define the dilution factor $\Delta$ by 
\beq
 \Delta \equiv 
 \frac{s_{\rm after}}{s_{\rm before}} =  \Max \lkk \left. \frac{\rho_M}{\rho_{\rm tot}} \right\vert_{T = T_{\rm ann}} , \ 1 \rkk, 
\eeq
When the dilution factor is larger than unity, 
the axion abundance is diluted by a factor of $\Delta^{-1}$ at $T = T_{\rm ann}$.

In the case that 
monopoles disappear after the QCD phase transition, 
the adiabatic suppression mechanism works around the QCD phase transition, 
so that the axion abundance is determined by \eq{Omegaa2} 
with a dilution factor: 
\beq
 \Omega_a h^2 
 \simeq 
0.4 \, \Delta^{-1} \frac{\theta_{\rm ini, H}^2}{\alpha_H^{2}} 
\lmk \frac{N_{\rm DW}}{N_{a,H}} \rmk
\lrf{10^{8}}{\Omega_M h^2} \lrfp{f_a}{\GEV{16}}{3}. 
\eeq
Note that in this case, the monopole abundance can be larger than the DM abundance
so that the condition of \eq{fa bound} can be avoided. 
In fact, the axion with a decay constant of order the GUT scale 
can be consistent with the observed DM abundance 
without a fine-tuning of axion misalignment angle.

The isocurvature constraint is the same as \eq{isocurvature constraint2} with $\Omega_a = \Omega_{\rm DM}$.
It can be rewritten as 
\beq
 H_{\rm inf} \lesssim 3.0 \times 10^{12} \GeV 
 \ \theta_{\rm ini, H} 
 \lrf{f_{a, H}}{\GEV{16}}.
\eeq

\section{Discussion and conclusions
\label{sec:conclusion}}

The adiabatic suppression mechanism is a novel mechanism to solve the moduli problem~\cite{Linde:1996cx}.  For the mechanism to work, the moduli must have a time-dependent mass much larger than the Hubble parameter.
In this case, the moduli follow the time-dependent minimum adiabatically, and thus the resultant oscillation amplitude is exponentially suppressed if one neglects the initial abundance generated by the effect of the inflaton oscillations~\cite{Nakayama:2011wqa,Nakayama:2011zy}.

In this paper, we have provided an analytic method to calculate the resultant adiabatic invariant which describes the comoving number density of homogeneous scalar field in the adiabatic suppression mechanism.
In particular, we have seen that the exponential suppression comes from a pole of the mass of the scalar field in the complex plane. 
The parameter dependence of our results are consistent with the result of Ref.~\cite{Linde:1996cx}, and moreover, 
our method can be used to calculate the number density in more generic models. 

Then we apply the result to axions. We have considered a model in which the axion obtains an effective mass due to the Witten effect in the presence of monopoles, and as a result, it starts to oscillate much before the epoch of QCD phase transition. This implies that the energy density of axion coherent oscillation can be much smaller than the case without the early oscillation. We have found that 
the axion energy density can be consistent with the observed DM abundance even in the case that the axion decay constant is 
as large as the GUT scale. Due to the early oscillation, 
both the axion energy density and its fluctuations are suppressed, 
so that the isocurvature problem can be ameliorated.

If U(1)$_H$ symmetry is not broken, 
the monopole is stable and also contribute to DM. 
Since monopoles interact with themselves via U(1)$_H$ gauge interaction,  they have a sizable velocity-dependent self-interaction cross section.  Such a self-interacting DM may be observationally preferred since it can relax the tension between the observed DM density profile and the prediction of $\Lambda$CDM model. 
Note however that, if stable, the monopole abundance has to be smaller than or equal to the observed DM density, 
so that the Witten effect on the axion is limited in this case. 
Thus we also considered the case that U(1)$_H$ symmetry is spontaneously broken at an intermediate scale so that the monopoles annihilate due to the tension of cosmic strings 
associated with U(1)$_H$ breaking. In this case, 
the monopole abundance in the early Universe is not related to the DM abundance, and so, the Witten effect can be much more significant. In particular, 
the axion density can be sufficiently suppressed even if its decay constant is as large as GUT scale.

\vspace{1cm}

%
\section*{Acknowledgments}
F.T. thanks Ken'ichi Saikawa for discussions. 
This work is supported by MEXT KAKENHI Grant Numbers 15H05889 (M.K. and F.T.), JP15K21733 (F.T.), and
JSPS KAKENHI Grant Numbers 17K05434(M.K.), 17H01131(M.K.), JP17H02875 (F.T.), JP17H02878(F.T.),  JP26247042(F.T.), and JP26287039 (F.T.), 
JSPS Research Fellowships for Young Scientists (M.Y.), 
and World Premier International Research Center Initiative (WPI Initiative), MEXT, Japan. 
%

\vspace{1cm}

\appendix{}

\section{The Witten effect on the QCD axion 
\label{sec:witten}}

In this appendix we explain the Witten effect on the axion 
and calculate its effective mass in the presence of monopole. 

We introduce a hidden Abelian gauge symmetry U(1)$_H$ with a monopole with mass of $m_g$ 
and a magnetic charge $g$. 
The Lagrangian of gauge fields is written as 
\beq
\mathcal{L} = 
 -  \frac{1}{4} F_{\mu \nu} F^{\mu \nu} - \frac{e^2 \theta }{32 \pi^2} 
 F_{\mu \nu} \tilde{F}^{\mu \nu}, 
\eeq
where $\tilde{F}^{\mu \nu} \equiv 1/2 \epsilon^{\mu \nu \sigma \rho} F_{\sigma \rho}$ 
and 
$e$ is the gauge coupling constant of the hidden gauge theory. 
In the presence of monopoles, the Maxwell's equations are given by 
\beq
 \del_\mu \tilde{F}^{\mu \nu} &=& j_M^\nu
 \\
 \del_\mu \lkk F^{\mu \nu} + \frac{e^2 \theta}{8 \pi^2} \tilde{F}^{\mu \nu}
 \rkk 
 &=& 0, 
\eeq
where $j_M^\nu$ is a monopole current. 
In the presence of monopoles, i.e., in the case of $j_M^\nu \ne 0$, 
there is no electromagnetic potential $A_\mu$ defined in the whole region. 
We can define it only in those regions where $j_M^\nu = 0$. 
The topology of these regions is nontrivial and in fact 
$A_\mu$ is singular in the presence of monopoles. 
This implies that the $\theta$-term in the Abelian gauge theory 
cannot be eliminated by integrating by parts 
in the presence of monopoles. 
As we see below, the $\theta$-term is physical and leads to an effect known as the Witten effect~\cite{Witten:1979ey}. 

The Gauss's laws are now modified as 
\beq
 \nabla \cdot {\bm B} &=& j_M^0, 
 \label{gauss1}
 \\
 \nabla \cdot {\bm E} + \frac{e^2}{8\pi^2} \nabla \cdot 
 \lmk \theta  {\bm B} \rmk 
 &=& 0, 
 \label{Gauss's law}
\eeq
where $E_i \equiv F_{0i}$ and $B_i \equiv -1/2\, \epsilon_{ijk} F^{jk}$. 
Let us emphasize that ${\bm E}$ and ${\bm B}$ are of the hidden electric and magnetic fields. 
We rewrite $j^0_M$ as 
\beq
 j_M^0 = g (n_{M^+} - n_{M^-}), 
\eeq
where $n_{M+} \,(n_{M^-})$ is the number density of (anti)monopoles. 
Then, Eq.~(\ref{Gauss's law}) implies that the monopole also carries an electric charge $q$,
which is proportional to $\theta$: 
\beq
 q = - \frac{e^2 g}{8 \pi^2} \theta. 
\eeq
This implies that the usual charge quantization condition, $q/e = n$, is extended to
\beq
 \frac{q}{e} + \frac{e g}{8 \pi^2} \theta = n, 
 \label{Witten effect}
\eeq
where $n$ is an integer that is nonzero for dyons. 
Therefore, the monopole is a dyon in a theory with a nonzero value of $\theta$, 
which is known as the Witten effect. 
Note that when we replace $\theta \to \theta + 2 \pi$, 
the value of $n$ changes from $n$ to $n + 1$ 
due to the Dirac's quantization condition: $g = 4\pi/e$.%
\footnote{
In our convention, 
half-integer electric charges are allowed. 
}
This means that the periodicity of $\theta$ is modified 
in the presence of monopole (dyon) such as $\theta \to \theta + 2 \pi$ and $n \to n +1$.

Now we assume that PQ symmetry is anomalous in terms of U(1)$_H$ and 
the QCD axion $a$ couples with U(1)$_H$. 
Thus we promote the theta angle of the hidden U(1)$_H$ theory 
to an axion by the PQ mechanism such as $\theta \to a/f_{a, H} - \theta$~\cite{Fischler:1983sc}: 
\beq
{\cal L}_\theta = - \frac{e^2  }{32 \pi^2} \lmk \frac{a }{f_{a, H}} - \theta \rmk
 F_{\mu \nu} \tilde{F}^{\mu \nu}.
 \label{aFF}
\eeq
where $f_{a, H}$ 
is the axion decay constant associated with U(1)$_H$. 
Suppose that $a$ and $a + 2\pi n f_{a,H}$ are physically identical where $n$ is a positive integer. 
Then, the smallest $n$ is called the domain wall number, $N_{a,H}$. 
Let us first consider a monopole located at the origin of a coordinate 
and calculate the effect of the axion on the energy density of electromagnetic field. 
The Maxwell's equation of \eq{gauss1} can be of course solved such as 
\beq
 {\bm B} = \frac{g}{4 \pi} \frac{\hat{r}}{r^2}, 
\eeq
where $\hat{r}$ is a normal vector along the radial direction. 
Together with \eq{Gauss's law}, 
this implies that 
the electric field is given by 
\beq
 {\bm E} = - \frac{e}{16 \pi^2} \frac{e g}{2 \pi} \lmk \frac{a (r) }{f_{a, H}} - \theta \rmk \frac{\hat{r}}{r^2}, 
\eeq
where $a( r = 0) = \theta f_{a, H}$. 
This means that a nonzero axion field value carries a large cost in the 
electrostatic field energy: 
\beq
 V &=& \int \dd^3 r \lkk \frac{1}{2} \lmk \nabla a \rmk^2 + \frac{1}{2} {\bm E}^2 \rkk, 
 \\
 &=& \int 4 \pi r^2 \dd r 
 \lkk \frac{1}{2} \lmk \frac{\del a}{\del r} \rmk^2 + \frac{e^2}{512 \pi^4} \lmk \frac{a }{f_{a, H}} - \theta \rmk^2 
 \frac{1}{r^4} \rkk, 
 \label{V_total}
\eeq
where we use $e g = 2 \pi$ in the second line.

We obtain the following axion configuration that minimizes the total energy of Eq.~(\ref{V_total}): 
\beq
 a(r) &=& a_0 \exp \lmk - r_0 / r \rmk + \theta f_{a, H}, 
 \\
 r_0 &=& \frac{e}{16 \pi^2 f_{a, H}}, 
\eeq
where $(a_0+\theta f_{a, H})$ denotes the asymptotic field value of the axion. 
The total energy is then given by 
\beq
 V_M 
 &=& 
 \beta f_{a, H} \lmk \frac{a }{f_{a, H}} - \theta \rmk^2, 
 \label{V_M}
 \\
 \left. \beta\right\vert_{\rm w/o \ cutoff}  
 &=& \frac{e}{8 \pi}, 
 \label{beta-wo cutoff}
\eeq
where we integrate Eq.~(\ref{V_total}) from $r = 0$ to $\infty$. 
However, the integral of \eq{V_total} may have to be taken in the interval of $[r_c, \infty)$ 
where $r_c$ is a cutoff scale due to, e.g., an electric screening effect 
of an electrically charged particle. 
This is actually true for a U(1)$_H$ gauge theory with a 't Hooft-Polyakov monopole, 
where there are charged massive gauge bosons after the SSB of SU(2) gauge symmetry. 
In that case, $r_c$ is given by the inverse of the mass of the charged gauge bosons. 
Then the parameter $\beta$ is given by 
\beq
 \left. \beta\right\vert_{\rm with \ cutoff}  
 &=& \frac{\alpha_H}{32 \pi^2} \frac{1}{r_c f_{a,H}}, 
 \label{beta_cutoff}
\eeq
where $\alpha_H \equiv e^2 / 4\pi$.

Note that the resulting energy \eq{V_M} is positive whether the monopole charge is positive or negative. 
As a result, 
the energy density of the axion ground state 
in a plasma with monopoles and antimonopoles  is given by $U = n_M V_0$, 
where 
$n_M = n_{M^+}+n_{M^-}$. 
\footnote{
We define $n_M$ as the sum of the number densities of monopoles and anti-monopoles, 
so that the axion mass squared is different from the one in Ref.~\cite{Fischler:1983sc} by a factor two. 
}
This implies that the axion obtains an effective mass of 
\beq
 \mam (T) = 2 \beta \frac{n_M (T)}{f_{a,H}}, 
 \label{FP effect}
\eeq
and has a VEV of $\theta$ at the minimum of the potential 
in a plasma with monopoles due to the Witten effect. 
Here we explicitly write the temperature dependence of $n_M$ due to, say, 
the expansion of the Universe.

Note that the axion VEV is determined such that 
the electric charge of monopole (dyon) is absent. 
As we can see from \eq{Witten effect}, 
the periodicity of $\theta$ is absent if we fix the value of $n$. 
This may imply that 
the axion does not have periodic potential but has a mass term of \eq{V_M} 
due to the Witten effect. 
However, 
monopoles (dyons) can have excited states 
with electric charges equal to or more than unity, 
which are known as Julia-Zee dyons~\cite{Julia:1975ff}. 
Such excited states can be produced at the monopole production process 
if, e.g., the PQ symmetry is broken after inflation 
and cosmic strings form at the SSB. 
In this case, 
the effective $\theta$ parameter (or axion VEV) 
changes $2 \pi N_{a, H}$ around the cosmic strings, 
where $N_{a, H}$ is the domain wall number. 
Then, at the monopole production process 
the Julia-Zee dyons with charge $n$ ($\le N_{a,H}-1$) may form 
in the domain of $\theta \in (2 \pi n - \pi, 2 \pi n + \pi)$. 
As a result, 
the domain wall may form at the boundary of nearby domains due to the Witten effect. 
In this paper, we do not consider this case and focus on the case where the PQ symmetry 
is broken before inflation.



\end{document}